\documentclass[aps,pra,twocolumn]{revtex4-1}

\usepackage{graphicx}
\usepackage{bm}                 % for bold math symbols
\usepackage{amsmath}            % for \text and such
\usepackage{amssymb}
\usepackage{verbatim}           % useful for commenting out stuff
\usepackage{amsthm}             % does theorems etc. nicely, but
\def\bra#1{{\langle#1|}}
\def\ket#1{{|#1\rangle}}
\def\bracket#1#2{{\langle#1|#2\rangle}}

\def\expect#1{{\langle#1\rangle}}

\def\tr{{\rm Tr}}

\def\Ahat{{\hat A}}

\def\Ehat{{\hat E}}

\def\Shat{{\hat S}}

\def\Zhat{{\hat Z}}

\def\Op{{\hat O}}
\def\id{{\hat I}}

\def\ep{\epsilon}

\def\la{\lambda}

\def\Eoo{\hat{E}_{00}}
\def\Eoi{\hat{E}_{01}}
\def\Eio{\hat{E}_{10}}
\def\Eii{\hat{E}_{11}}
\def\Mjk{\hat{M}_{jk}}
\def\Rhat{\hat{R}}
\def\Qhat{\hat{Q}}

\newtheorem{theo}{Theorem}
\newtheorem{dfn}{Definition}

\def\beq{\begin {equation}}
\def\eeq{\end {equation}}
\def\dd#1#2{{\partial #1 \over \partial #2}}
\def\lb{ \left[ }
\def\rb{ \right]  }
\begin{document}

\title{Compatibility of state assignments and pooling of information}

\author{Todd A. Brun}
\email{tbrun@usc.edu}
\affiliation{Communication Sciences Institute, University of Southern California,
Los Angeles, California  90089, USA }
\author{Min-Hsiu Hsieh}
\email{min-hsiu.hsieh@uts.edu.au}
\affiliation{Centre for Quantum Computation \& Intelligent Systems (QCIS),
Faculty of Engineering and Information Technology,
University of Technology, Sydney, NSW 2007, Australia}
\author{Christopher Perry}
\email{christopher.perry.12@ucl.ac.uk }
\affiliation{Department of Physics and Astronomy, University College London,
Gower Street, London WC1E 6BT, United Kingdom}

%%% abstract before title in revtex4 %%%
\begin{abstract}
We say that two (or more) state assignments for one and the same quantum system are {\it compatible} if they could represent the assignments of observers with differing information about the system.  A criterion for compatibility was proposed in [\textit{Phys.~Rev.~A 65, 032315 (2002)}]; however, this leaves unanswered the question of whether there are {\it degrees} of compatibility which could be represented by some quantitative measure, and whether there is a straightforward procedure whereby the observers can pool their information to arrive at a unique joint state assignment.  We argue that such measures are only sensible given some assumption about what kind of information was used in making the state assignments in the first place, and that in general state assignments do not represent all of the information possessed by the observers.  However, we examine one particular measure, and show that it has a straightforward interpretation, assuming that the information was acquired from a particular type of measurement, and that in this case there is a natural rule for pooling information.  We extend this measure to compatibility of states for $k$ observers, and show  that the value is the solution to a semidefinite program.  Similar compatibility measures can be defined for alternative notions of state compatibility, including {\it Post-Peierls} (PP) and {\it Equal Support} (ES) compatibility.
\end{abstract}

\date{\today}

\maketitle

\section{Introduction}

While there has been much debate about the exact nature of the quantum state, the most persuasive interpretation is that the state represents the knowledge or  belief of an observer about a given quantum system.  Using the state, the observer can assign probabilities to the outcomes of any possible measurement of the system.  From this point of view, a pure state represents a state of maximal knowledge (or minimal ignorance) about a system, and a mixed state represents a description with incomplete knowledge.  Whether the state also reflects the structure of an underlying physical reality is a subject of heated debate; but at a practical level, we make state assignments based on our available information.

If the state assignment reflects the information of an observer, it follows reasonably that two observers with different information will make different state assignments.  This can arise naturally in a number of ways; for example, one observer may have access to certain measurement outcomes that another observer does not, or may have additional information about how the system was prepared.  In \cite{Brun:2002eo} this question was examined, and a necessary and sufficient condition was found which must be satisfied by any two or more state assignments which reflect differing information about the same physical system, provided that the information underlying all of these assignments is accurate and reliable.

Classically, the state of knowledge about a system is described by a probability distribution.  New information about the system can be acquired by performing a measurement, and updating the state assignment based on a measurement result is done using the Bayes rule.  Classically, two probability distributions can be compatible so long as they are not actually contradictory.  Furthermore, it is possible to pool the information of different observers using only their individual states of knowledge (probability distributions), {\it provided} that the information of different observers is independent.  In this case, other information (such as how the knowledge was obtained) is not required.

A reasonable question is whether similar conditions hold in quantum mechanics.  Can we take the state assignment to reflect all information about a quantum system?  If so, under what conditions?  What assumptions do we have to make about how the information underlying the state assignments was acquired?

In this paper, we look at different ways that a state can be derived from underlying information, and show that in general the state assignment does not completely summarize all the information used to derive it.  In particular, we consider two significantly different ways of constructing the state.  In the first case, the information is derived from measurement. Each party obtains his or her information by making a measurement. In the second case, each observer is given {\it classical} information about how the system was prepared, for example from the third party who prepared it.  The method of pooling information becomes clear once we know how the information was obtained, but different types of information can result in very different ways of  acquiring a joint state assignment. However, this does not rule out the possibility that within a very sharply defined context, it may be possible and useful to both define reasonable measures of compatibility, and give rules for for pooling state assignments.

There have been various studies concerning compatibility of state assignments and pooling of information \cite{Brun:2002eo,Jacobs:2002cc,Caves:2002ec,Poulin:2003he,Herbut:2004vc,Jacobs:829374,Spekkens:2007wn,Leifer:2011ug}. Quite recently, the authors in Ref.~\cite{Leifer:2011ug} used the conditional states formalism \cite{Leifer:2011wq} to show that earlier results \cite{Jacobs:2002cc,Poulin:2003he,Herbut:2004vc,Jacobs:829374,Spekkens:2007wn} can be recast in this more general formalism. Our results differs from theirs in that the main focus of this paper is an attempt to \emph{quantify} the degree of compatibility in the assigned quantum states. We also note that the problem of pooling information, so that a unique joint state assignment can be derived, is very sensitive to the type of information used to make the state assignment in the first place, and illustrate this with a couple of representative examples.

In Sec.~\ref{sec2}, we briefly describe the necessary and sufficient condition for two different states to be compatible. We discuss the possible existence of a compatibility measure (or measures) that quantifies how much the two states are compatible. In Sec.~\ref{sec3}, we show two significantly different ways of obtaining the state assignment, and argue that the state does not encapsulate the whole information in general.  In Sec.~\ref{SMR}, we explore a particular compatibility measure, and show that this measure is actually a distance measure between two states, while in Sec.~\ref{sec5} we show how if we assume that the state assignments were derived from a particular type of measurement, there is a simple rule for forming a joint state assignment.  In Sec.~\ref{SDP_sec} we extend this measure to compatibility of states for $k$ observers, and show that it is the solution of a semidefinite program; we then define similar compatibility measures for two more restrictive notions of compatibility, Post-Peierls (PP) and Equal Support (ES) compatibility.  We conclude the paper in Sec.~\ref{sec7}.

\subsection{Information, Knowledge, and Belief}

It seems necessary to make a bit more concrete what we mean when we talk about different observers have different {\it information} about a given system.  What, exactly, do we mean by information in this context?

For clarity, we will draw somewhat arbitrary distinctions between {\it information}, {\it knowledge}, and {\it belief}.  When does it make sense even to  talk about compatibility and state pooling?  Arguably, two observers will only have a basis for comparison between their state assignments if they start from a fairly substantial base of shared knowledge about the system.

For example, the two observers might both know that the give system was prepared by a given type of experimental apparatus, or by performing a particular generalized measurement.  This would represent their shared knowledge base about the system.  Given that shared knowledge, the two observers might still have different {\it information} about the system.  For instance, they might each have partial, but different, information about the settings of the device that prepared the system; or they might each have partial, but different, information about the outcome of the generalized measurement.  When we refer to {\it information} in the discussion that follows, we will mostly be using it in this rather restrictive sense---that is, some amount of numerical data about the system, or more generally, a probability distribution over such data.

If the two observers do not have a shared knowledge base then it becomes harder to compare their state assignments, or for such comparisons to even make sense.  If two observers think that the system was prepared in radically different ways, then it is hard to see how they could share their knowledge, even if their state assignments were almost identical.  This is where knowledge shades over into belief.  A number of authors have pointed out that two observers in possession of the same facts might nevertheless arrive at quite different state assignments if they have different {\it prior distributions} or {\it priors}.  (See, for example, \cite{Caves:2002ec}.)

In most discussions of compatibility---including that in this paper---we are implicitly assuming that the observers with compatible state assignments begin with identical (or very similar) priors about the system, and a strong base of shared knowledge.  Their state assignments differ due to the acquisition of different data about the system.  While there is certainly no necessity that different observers should start from the same prior---quite the contrary---in practice, it is quite common in science for observers to come to consensus about particular experimental systems, generally by sharing large amounts of data from repeated trials or preparations.

It would be interesting to study the idea of compatibility for observers with different priors, and in particular the process whereby consensus can be reached, but that is beyond the scope of the current paper.

\section{Compatible state assignments}
\label{sec2}

Let $\rho_A$ and $\rho_B$ be two different assignments of density matrix to the same system.  If we assume that these assignments were made by two observers with different information, what restriction does this place on the state assignments?  We call two states that satisfy such a restriction {\it compatible}.  A necessary and sufficient criterion \cite{Brun:2002eo} for $\rho_A$ and $\rho_B$ to be
compatible is that {\it the intersection of their supports} is nonempty:
\begin{equation}
\text{supp}(\rho_A)\cap \text{supp}(\rho_B)\neq\emptyset.
\label{compatibility}
\end{equation}
To show that this condition is necessary, we assume that the information used to derive state descriptions $\rho_A$ and $\rho_B$ is {\it accurate and reliable}.  The two observers should then be able to combine their information to produce a joint state description $\rho_{AB}$.  Since their information is accurate and reliable, any measurement result which is ruled out (i.e., assigned zero probability) by either party should also be ruled out in the joint state $\rho_{AB}$.  This then implies
\beq
\begin{split}
\text{null}(\rho_A) , & \text{null}(\rho_B) \subseteq \text{null}(\rho_{AB})  \\
& \Rightarrow \text{span}\{\text{null}(\rho_A),\text{null}(\rho_B)\}\subseteq \text{null}(\rho_{AB}) \\
&\Rightarrow \text{supp}(\rho_{AB})\subseteq \text{supp}(\rho_{A})\cap\text{supp}(\rho_{B}) .
\end{split}
\eeq
Sufficiency is proven by constructing an explicit situation in which $\rho_A$ and $\rho_B$ could arise as state estimates by two observers making different measurements.  We introduce purifications of $\rho_A$ and $\rho_B$ with two ancillas, and then combine them into a single state. The given state then results from Alice and Bob each measuring his or her own ancilla and getting a particular outcome.  See \cite{Brun:2002eo} for details.

The compatibility criterion in Eq.~(\ref{compatibility}) is robust against sufficiently small distortions of the states. For example, given $\rho_A=\ket{0}\bra{0}$ and  $\rho_B=\ep\ket{0}\bra{0}+(1-\ep)\ket{1}\bra{1}$, the two states remain compatible until $\ep\to 0$.  Unfortunately, it is in some ways not very informative.  The criterion is an all-or-nothing property; it says only that two state assignments {\it could} result from observers with differing information, but gives no clue how likely it is that they did.  This makes it natural to seek a measure of compatibility that would indicate this likelihood. The measure would be zero for incompatible states, and go up to one for identical states.  We discuss the possibility of such measures in Secs.~\ref{SMR} and \ref{SDP_sec}.

We note that other (more restrictive) notions of compatibility also exist:  see Ref.~\cite{Caves:2002ec} for an extensive discussion of this subject and a variety of different compatibility criteria.  We briefly look at two of these other criteria in Sec.~\ref{SDP_sec}.

\section{Types of information}
\label{sec3}

We think of the state assignment as being based on information possessed by the observer; but information can come in many different forms.  In this paper, we consider two different types of information in particular:  classical information about the state preparation, in which the state assignment is made using the {\it maximum entropy estimate}, and information derived from measurements.

\subsection{Maximum entropy estimate}

If an observer acquires classical information about the way a system was prepared, he or she can make a state assignment based on the {\it maximum entropy principle} \cite{Jaynes}:  choosing the state with the highest entropy consistent with the given information. This takes a particularly simple form if the information is in the form of an expectation value $\langle \Op \rangle \equiv \tr \{\Op\rho\}=o$.  The observer can construct a state assignment by maximizing the Von~Neumann entropy
\begin{equation}
S(\rho)=-\tr\{\rho\log\rho\} ,
\label{vonNeumann}
\end{equation}
under the following two constraints:
\beq
\begin{split}
\label{constrain}
f(\rho) \equiv &\tr \{\Op\rho\} = o\\
g(\rho) \equiv &\tr\{\rho\} = 1.
\end{split}
\eeq
This is a constrained maximization problem, and can be solved with Lagrange multipliers:
\begin{equation}
\begin{split}
\dd{S(\rho)}{\rho} + \lambda_1 \dd{f(\rho)}{\rho}
  + \lambda_2 \dd{g(\rho)}{\rho} = 0 \\
\Rightarrow  \log\rho+\id = \lambda_1 \Op+\lambda_2\id .
\end{split}
\end{equation}
The estimated state is
\begin{equation}
\tilde{\rho} = \frac{e^{\lambda_1\Op+(\lambda_2-1)\id}}{\tr\{e^{\lambda_1\Op+(\lambda_2-1)\id}\}} ,
\end{equation}
where we can solve for $\lambda_1$ and $\lambda_2$ from (\ref{constrain}). This assignment is obviously based on the observer's knowledge; further information can alter the assignment. If an additional piece of information $\langle \Ahat \rangle=a$ is given, the state changes accordingly:
\begin{equation}
\tilde{\rho}' =
  \frac{e^{\la_1'\Op+\la_2'\Ahat+(\la_3'-1)\id}}{\tr\{e^{\la_1'\Op+\la_2'\Ahat+(\la_3'-1)\id}\}}.
\end{equation}
If Alice and Bob wish to form a joint state assignment, they must share all the information they have about the state preparation and perform a maximum entropy assignment with this shared information.  It is easy to see that as long as all the information provided to Alice and Bob is consistent---that is, that there exists a state with all those expectation values---their state assignments will be compatible.  Moreover, the entropy of the joint state $\rho_{AB}$ must be less than or equal to the entropies of the individual states:  $S(\rho_{AB}) \le S(\rho_A)$ and $S(\rho_{AB}) \le S(\rho_B)$.

\subsection{Learning measurement results}

Consider now a different situation.  Suppose that Alice and Bob share two halves of an entangled state, which we write in the Schmidt decomposition:
\[
\ket\Psi=\sum_{j}\la_j\ket{\psi_j}\otimes\ket{\phi_j} ,
\]
with $\langle \psi_j|\psi_k\rangle=\langle\phi_j|\phi_k\rangle=\delta_{jk}$.  Alice has the reduced state
\begin{equation}
\rho_A=\tr_{B}\{\ket{\Psi}\bra{\Psi}\}=\sum_{j}\la_j^2\ket{\psi_j}\bra{\psi_j} ,
\label{reduced_Alice}
\end{equation}
which can accurately predict the probabilities for any measurement she makes.  Suppose Bob makes a measurement in the basis $\{\phi_j\}$, and gets outcome $k$.  Then Alice's state should immediately become $\rho_A=\ket{\psi_k}\bra{\psi_k}$.  She cannot, however, update her state until she learns the outcome of Bob's measurement.  Until that point, her original state assignment $\rho_A$ is the best  she can make.

\subsection{The state does not include all information about the system}

As mentioned in Sec. 1, the state itself does not always include all the information an observer has about a system.  For information acquired from measurements, this was shown by Jacobs \cite{Jacobs:2002cc}, who demonstrated that if $\rho_A$ and $\rho_B$ are compatible state assignments, which are assumed to be obtained from the results of measurements by Alice and Bob, then it is possible to construct an initial state and choice of measurement such that the joint state assignment $\rho_{AB}$ can be {\it any} density matrix at all, so long as $\text{supp}(\rho_{AB}) \subseteq \text{supp}(\rho_A) \cap \text{supp}(\rho_B)$.  Thus, in general more information is needed than $\rho_A$ and $\rho_B$ in order to construct $\rho_{AB}$; $\rho_A$ and $\rho_B$ do not encapsulate all of Alice and Bob's information about the system.

Here we give another example showing that the same is true if Alice and Bob make their state assignments using classical information and the maximum entropy principle.  Suppose Alice assigns the state of a qubit:
\begin{equation}
\rho_A = (\id+a\hat{X}+b\hat{Y}+c\hat{Z})/2 ,
\end{equation}
where $a^2+b^2+c^2\leq1$.  This same state can be obtained from classical information in many ways.  Here are two examples:
\beq
\begin{split}
\text{1.}&\ \langle a\hat{X}+b\hat{Y}\rangle=a^2+b^2,\ \langle \hat{Z} \rangle=c \\
\text{2.}&\ \langle b\hat{Y}+c\hat{Z}\rangle=b^2+c^2,\ \langle \hat{X} \rangle=a.
\end{split}
\eeq
Suppose that Bob makes his state assignment
\begin{equation}
\rho_B = (\id + d\hat{Y})/2
\end{equation}
based on the classical information that $\langle \hat{Y} \rangle=d$.  We can readily see that $\rho_A$ and $\rho_B$ are compatible.  (Indeed, any two qubit states are compatible unless they are two distinct pure state assignments.)  Now suppose that Alice and Bob share their information to obtain a joint state assignment $\rho_{AB}$.  The joint state in cases 1 and 2 will be different:
\beq
\begin{split}
\text{1.}&\  \rho_{AB} = (\id+(a+b(b-d)/a)\hat{X}+d\hat{Y}+c\hat{Z})/2 \\
\text{2.}&\  \rho_{AB} = (\id+a\hat{X}+d\hat{Y}+(c+b(b-d)/c)\hat{Z})/2.
\end{split}
\eeq
This shows that knowledge of the state assignments $\rho_A$ and $\rho_B$ alone is not sufficient to know how to pool information in the case where the information is classical, as well.  The state assignment does not encapsulate all information about the system.

\section{Measures of compatibility}
\label{SMR}

The compatibility criterion of \cite{Brun:2002eo} is all-or-nothing:  either two states are compatible, or they are not.  This includes extreme examples like $\rho_A = \ket0\bra0$ and $\rho_B = \epsilon \ket0\bra0 + (1-\epsilon)\ket1\bra1$.  For any $\epsilon > 0$ these states are compatible, even though they are practically orthogonal for very small $\epsilon$.  This leads to the natural question:  can we define {\it measures} of compatibility?  Intuitively, while compatibility indicates that two state assignments {\it could} represent different information about the same system, such a measure would represent a probability that they actually {\it do}.  We should therefore expect the measure to go from 0 for incompatible states to 1 for identical states, with some kind of smooth behavior in between.

From our earlier discussion, some caveats are clearly needed.  Since the states do not represent all information about the system, it is impossible to {\it truly} measure the compatibility based on the state assignments alone.  For example, it is impossible to arrive at compatible state assignments using classical information which is contradictory; e.g., the two pieces of classical information $\expect{\hat{X}} = 0.5$ and $\expect{\hat{X}} = -0.5$ lead to compatible state assignments for a qubit, but are obviously inconsistent with each other.  (This inconsistency comes from the assumptions of the  Maximum Entropy procedure, where these expected values are taken to be guaranteed properties of the state.)  Therefore, in defining a compatibility measure, one must implicitly assume that the state assignments represent information which was acquired in a consistent manner---for example, by Alice and Bob performing measurements on the same system.

One such measure was proposed by Poulin and Blume-Kohout, in close analogy with the measure of distance between classical probability distributions \cite{Poulin:2003he}.  The correct interpretation of this measure is not exactly clear in the quantum case, but it does have the right sort of qualitative behavior, and reduces to a known measure in the case of essentially classical state (i.e., when $\rho_A$ and $\rho_B$ commute).  In other cases, however, it is somewhat difficult to compute, since it requires taking a minimum over all pure state decompositions of density matrices.

Here we examine a different compatibility measure, which was originally suggested by Kitaev \cite{Kitaev}. The original idea is to find a positive matrix $\Rhat$ that ``fits'' into both the state matrix $\rho_A$ and $\rho_B$ to the greatest degree possible.
\begin{dfn}[Compatibility Measure]
Let $\rho_A$ and $\rho_B$ be compatible states. Consider all positive matrices $\Rhat \ge 0$ such that
\beq
\begin{split}
\label{MR}
\rho_A-\Rhat &\geq 0 , \\
\rho_B-\Rhat &\geq 0 .
\end{split}
\eeq
The compatibility measure $K$ is defined as
\begin{equation}
K(\rho_A,\rho_B)=\max_{\Rhat\geq 0}\tr \Rhat .
\label{KitaevMeasure}
\end{equation}
\end{dfn}
We have found no simple formula for this measure, in general.  (In Sec.~\ref{SDP_sec} we show that this value can be found as the solution of a semidefinite program.)  However, it is possible to upper bound $K(\rho_A,\rho_B)$ with the {\it trace distance} between $\rho_A$ and $\rho_B$ \cite{Nielsen}):
%\begin{theo}
\[
K(\rho_A,\rho_B) \le 1-D(\rho_A,\rho_B)
\]
where $D(\rho_A,\rho_B)$ is the trace distance:
\begin{equation}
D(\rho_A,\rho_B) \equiv \frac{1}{2}\tr\left\{ |\rho_A-\rho_B| \right\}.
\end{equation}
%\end{theo}
% {\bf Proof:} \,\,\,\,
To see this inequality, define $\tilde{R}=\rho_A-\Rhat$.  We can rewrite Eqs.~(\ref{MR}) and (\ref{KitaevMeasure}) in terms of $\tilde{R}$:
\beq
\begin{split}
\label{MR1}
\tilde{R} &\geq 0 , \\
\rho_B-\rho_A+\tilde{R} &\geq 0 , \\
\rho_A - \tilde{R} &\geq 0 ,
\end{split}
\eeq
and
\begin{equation}
K(\rho_A,\rho_B)=\max_{\Rhat\geq 0}\tr\{\Rhat\} = 1 - \min_{\tilde{R}\geq 0}\tr\{\tilde{R}\} .
\label{bound}
\end{equation}
This just rewrites the original definition of the measure in terms of the new operator $\tilde{R}$.  However, we can put an upper bound on this quantity by relaxing the third requirement above that $\rho_A - \tilde{R} \geq 0$.  In this case, we can find an exact expression for (\ref{bound}).  From \cite{NielsenChuang}, we know that $\rho_B-\rho_A$ can be expressed as
\begin{equation}
\rho_B-\rho_A=\Qhat-\Shat,
\end{equation}
where $\Qhat$ and $\Shat$ are positive operators with orthogonal support.  This implies that $|\rho_B-\rho_A|=\Qhat+\Shat$.  We can satisfy (\ref{MR1}) and minimize $\tr\{\tilde{R}\}$ by choosing \begin{equation}
\tilde{R}=\Shat=\frac{1}{2}(|\rho_B-\rho_A|-\rho_B+\rho_A).
\end{equation}
Then
\begin{equation}
K(\rho_A,\rho_B) \le 1-\frac{1}{2}\tr\left\{ |\rho_B-\rho_A| \right\} = 1 - D(\rho_A,\rho_B) .
\end{equation}

It is easy to see that this measure has the right qualitative behavior for a measure of compatibility:  it goes from $0$ for incompatible states (where, in general, the above bound will be nonzero unless the states are actually orthogonal) to $1$ for identical states (at which point it will agree with the above bound).

\begin{figure}
    \centering
        \includegraphics[width=0.5\textwidth]{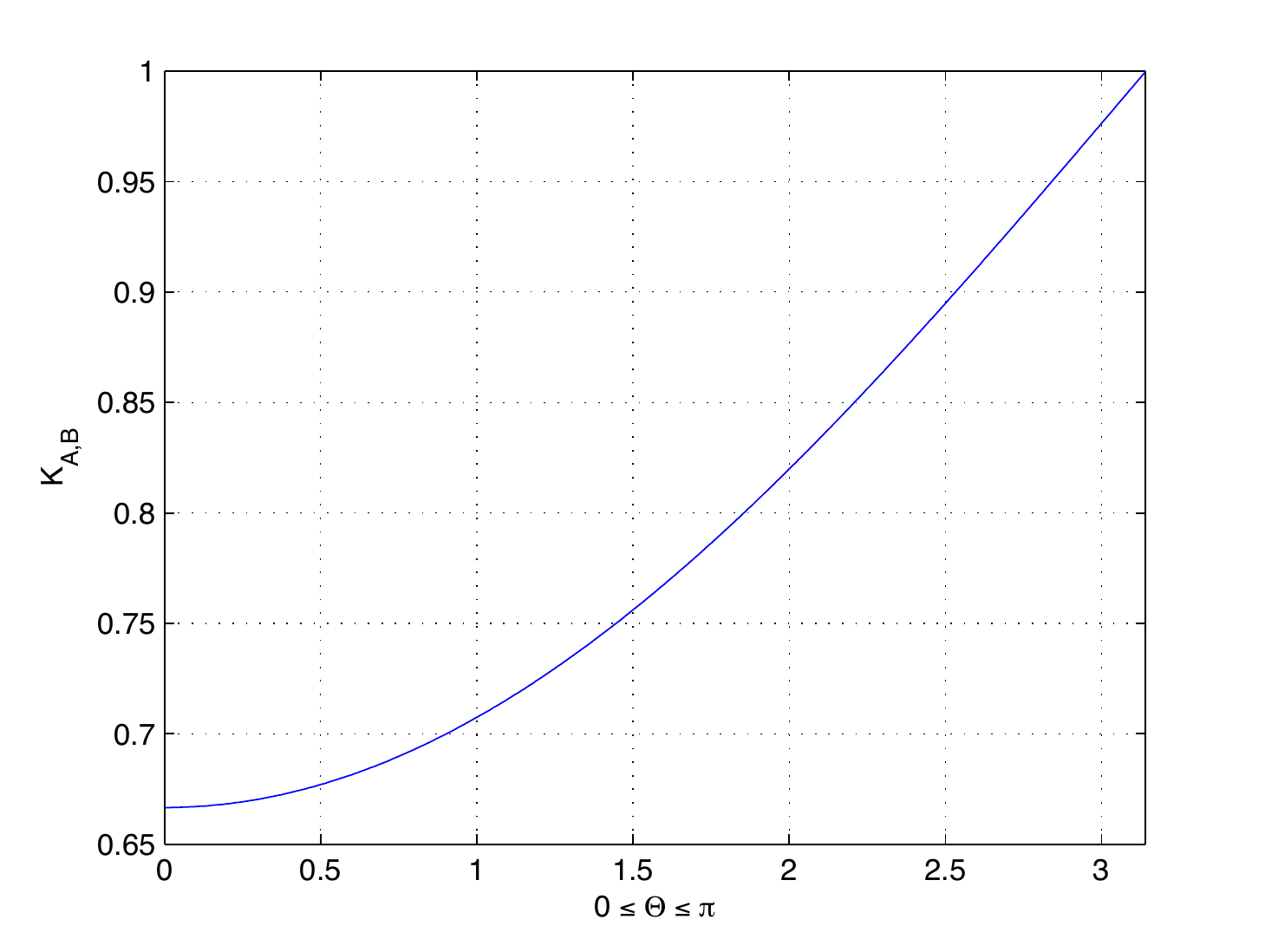}
    \caption{Compatibility measure between $K_{A,B}$ versus angle $\theta$}
    \label{fig:ME}
\end{figure}

We will now show a couple of examples of this compatibility measure for states of qubits, in which the information which produces the state assignment is either classical information about the preparation, or the results of measurements.  For our first example, suppose that Charlie prepares the pure state
\[
\rho=(\id+\vec{r}\cdot\hat{\vec{\sigma}})/2 ,
\]
where $\vec{r}=(r_x,r_y,r_z)$ and $\|\vec{r}\|=1$.  Define an Hermitian operator $\hat{O}=\cos(\theta/2)\Zhat+\sin(\theta/2)\hat{X}$.  Charlie gives the expectation values $o = \langle\hat{O}\rangle = r_x\sin(\theta/2)+r_z\cos(\theta/2)$ to Alice, 
and $x=\langle\hat{X}\rangle=r_x$ to Bob.  Alice and Bob's state estimates are, respectively,
\begin{equation}
\begin{split}
\rho_A &= (\id+o\cdot\hat{O})/2=(\id+\vec{r}_A\cdot\hat{\vec{\sigma}})/2 , \\
\rho_B &= (\id+x\cdot\hat{X})/2=(\id+\vec{r}_B\cdot\hat{\vec{\sigma}})/2 ,
\end{split}
\end{equation}
where $\vec{r}_A = (o\sin(\theta/2),0,o\cos(\theta/2))$ and $\vec{r}_B = (r_x,0,0)$.
The measure $K(\rho_A,\rho_B)$ is then
\beq
\begin{split}
K(\rho_A,\rho_B) & = 1-\frac{\|r_A-r_B\|}{2} \\
& = 1-\frac{1}{2}\cos(\theta/2)\sqrt{r_x^2+r_z^2} .
\end{split}
\eeq
This expression obviously depends both on Charlie's choice of state and his choice of observables to pass on to Alice and Bob.  If we average this over all possible pure states, then we get a simple result which depends only on the choice of observable:
\beq
\begin{split}
K_{A,B}&=\mathbf{E}_{\|\vec{r}\|=1}\lb 1-\frac{1}{2}\cos(\theta/2)\sqrt{r_x^2+r_z^2}\rb  \\
&=1 - (1/3)\cos(\theta/2) .
\end{split}
\eeq
We plot this in Fig.~\ref{fig:ME}.  The states $\rho_A$ and $\rho_B$ become identical when $\theta=\pi$, and therefore are perfectly compatible.

\begin{figure}[t]
    \centering
\includegraphics[width=0.5\textwidth]{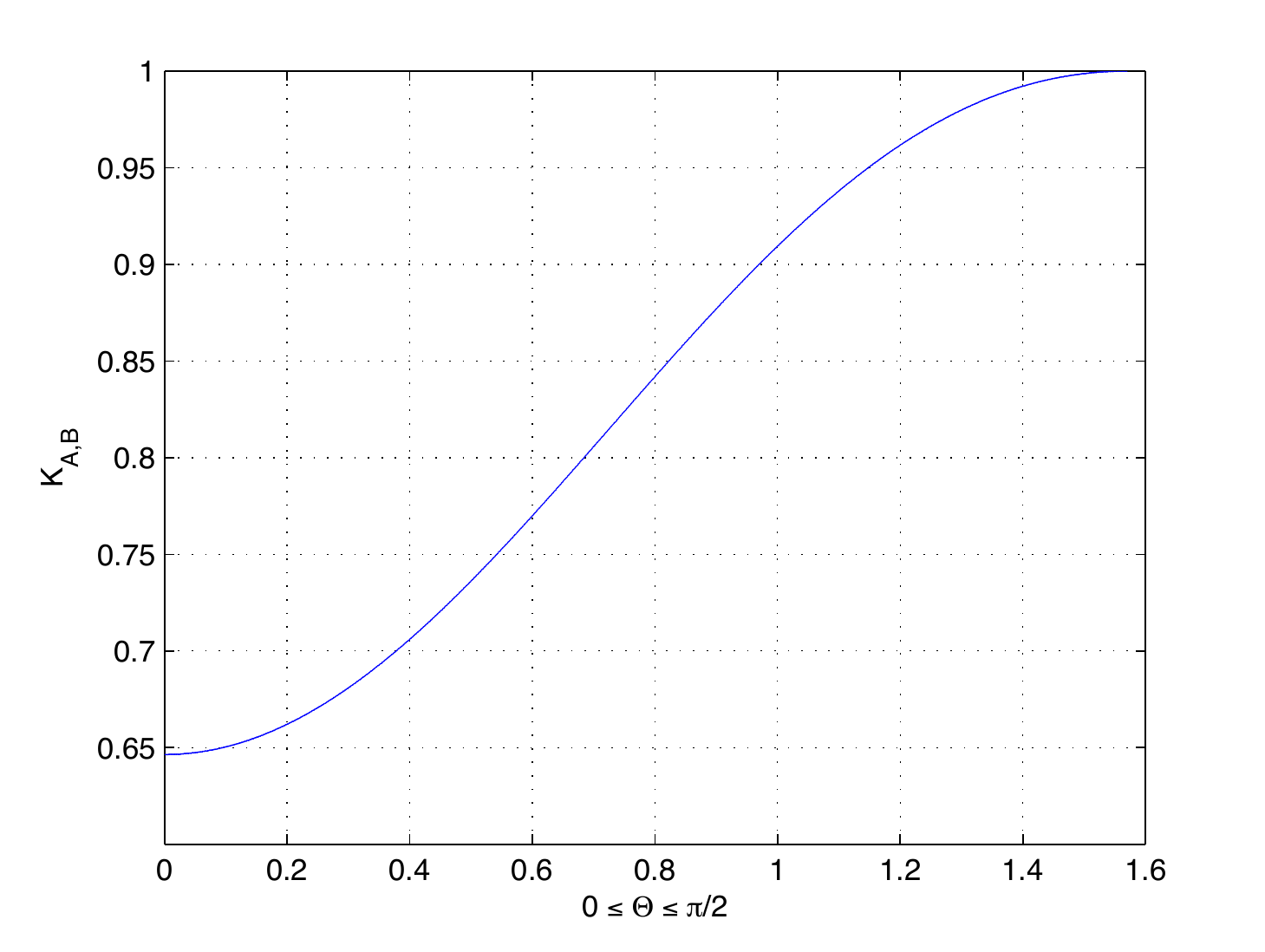}
\caption{Compatibility measure $K_{A,B}$ versus angle $\theta$}
\label{fig:MM.eps}
\end{figure}

In our second example, the information comes from measurement.  Assume that the initial state is unknown to Alice and Bob, so that they assume a completely mixed state.  That is,
\[
\rho_0 = \id/2 .
\]
Alice performs projective measurement $\{A_0,A_1\}$ with outcome $0$ and $1$,
\beq
A_0=\ket{\Psi_0}\bra{\Psi_0} , \ \ \ 
A_1=\ket{\Psi_1}\bra{\Psi_1} ,
\eeq
where $\ket{\Psi_0}=\cos\frac{\theta}{2}\ket{0}+\sin\frac{\theta}{2}\ket{1}$ and
$\ket{\Psi_1}=-\sin\frac{\theta}{2}\ket{0}+\cos\frac{\theta}{2}\ket{1}$.  For $\theta=0$ this is a $\hat{Z}$ measurement; for $\theta=\pi/2$ it is an $\hat{X}$ measurement.  Bob also performs a projective measurement of $\hat{X}$:
\beq
B_0=\ket{+}\bra{+} , \ \ \ 
B_1=\ket{-}\bra{-} .
\eeq
However, as in the paper of Jacobs \cite{Jacobs:829374}, we assume that Alice and Bob don't know the order of their measurements, and that they each know only the outcomes of their own measurements.  Their state estimates are thus
\beq
\begin{split}
\rho_A^i=\sum_{j=0}^{1}\frac{B_jA_iA_i^{\dagger}B_j^{\dagger}+A_iB_jB_j^{\dagger}A_i^{\dagger}}{p_{i}},\
p_i=\sum_{j=0}^{1}p_{ij} \\
\rho_B^j=\sum_{i=0}^{1}\frac{B_jA_iA_i^{\dagger}B_j^{\dagger}+A_iB_jB_j^{\dagger}A_i^{\dagger}}{p_{j}},\
p_j=\sum_{i=0}^{1}p_{ij} \\
\end{split}
\eeq where
$p_{ij}=\tr\{B_jA_iA_i^{\dagger}B_j^{\dagger}+A_iB_jB_j^{\dagger}A_i^{\dagger}\}$.
Let $K_{i,j}=K(\rho^i_A,\rho^j_B)$, and $K_{A,B}=\mathbf{E}[K_{i,j}]=\sum_{i,j}p_{ij}K_{i,j}$.
Shown in Fig. (\ref{fig:MM.eps}) is the compatibility measure
$K_{A,B}$ versus the angle $\theta$,  
where the compatibility measure saturates when $\theta=\pi/2$. 

One caveat should be inserted here.  When the degree of compatibility is high, this implies that Alice and Bob each have information that leads to almost the same state assignment.  This might lead one to think that when the degree of compatibility is low, their information is only marginally consistent.  But this need not be the case.  Consider the example given in Eq.~(\ref{reduced_Alice}).  While Alice's and Bob's state assignments are guaranteed to be compatible, it is possible for their degree of compatibility to be quite low---particularly if Bob makes a measurement with many possible outcomes.  But their information is clearly perfectly consistent; Alice merely lacks one piece of data possessed by Bob.

As we will see in the next section, this compatibility measure has a natural interpretation in terms of a particular method for gathering information, which also leads to a natural rule for pooling states.

\section{Pooling states}
\label{sec5}

Kurt Jacobs showed \cite{Jacobs:2002cc} that one can construct different measurement situations---that is, initial states, choices of measurement, and measurement outcomes---which yield $\rho_A$ and $\rho_B$ when Alice and Bob each have only partial knowledge of the outcomes, but that upon sharing information will produce any state $\rho_{AB}$ at all, so long as the support of $\rho_{AB}$ is in the
intersection of the supports of $\rho_A$ and $\rho_B$. Given only $\rho_A$ and $\rho_B$, the problem of determining $\rho_{AB}$---information pooling---is not uniquely solvable.  However, if we make particular assumptions about the type of measurement used to determine $\rho_A$ and $\rho_B$, we can single out a particular state $\rho_{AB}$, by choosing the measurement which maximizes the probability of the outcome yielding $\rho_A$ and $\rho_B$.  This probability turns out to be essentially identical to the compatibility measure $K(\rho_A,\rho_B)$.

Assume that Alice and Bob have no information regarding the quantum system, then
the density matrix $\rho_0$ is maximally mixed,
\[
\rho_0 = \id/D ,
\]
where $D$ is the dimension of the state space.  Alice and Bob carry out some joint measurement $\{\Mjk\}$, where $\sum_{jk}\Mjk^{\dagger}\Mjk=I$, and Alice gets result $j$ while Bob gets result $k$.  Then
\begin{eqnarray}
\label{pool1}
\rho_A &=& \frac{\displaystyle \sum_{k}\Mjk \Mjk^{\dagger}}{\displaystyle \sum_{k}\tr\left\{ \Mjk^{\dagger}\Mjk \right\} } \nonumber\\
\rho_B &=& \frac{\displaystyle \sum_{j}\Mjk \Mjk^{\dagger}}{\displaystyle \sum_{j}\tr\left\{ \Mjk^{\dagger}\Mjk \right\}}   \nonumber\\
\rho_{AB} &=& \frac{\Mjk \Mjk^{\dagger}}{\tr\left\{\Mjk^{\dagger}\Mjk \right\}}
\end{eqnarray}
Given fixed states $\rho_A$ and $\rho_B$, consider all measurements which yield $\rho_A$ and $\rho_B$ for some pair of outcome of $j$ and $k$.  We will show that the particular measurement that maximizes the probability of $\rho_{AB}$ is the positive matrix that satisfies (\ref{MR}).  This gives a particular interpretation for the compatibility measure defined in section 4.

For simplicity, we start by considering only positive measurement operators,
\[
\Mjk^\dagger = \Mjk \ge 0,\ \ 
\Mjk^2 \equiv \Ehat_{jk} .
\]
We can lump together all the measurement outcomes other than the ``correct'' one (i.e., the one that is assumed to actually occur), so we can restrict ourselves to the case where $j=0,1$ and $k=0,1$, and
\[
\Eoo+\Eoi+\Eio+\Eii=\id
\]
Without loss of generality, we assume that the outcome $(0,0)$ is the case of interest. Then we can rewrite (\ref{pool1}) as
\beq
\begin{split}
\label{pool2}
\rho_A &=\frac{\Eoo+\Eoi}{\tr\{\Eoo+\Eoi\}} , \\
\rho_B &=\frac{\Eoo+\Eio}{\tr\{\Eoo+\Eio\}} , \\
\rho_{AB} &=\frac{\Eoo}{\tr\{\Eoo\}} .
\end{split}
\eeq

Let $a \equiv \tr(\Eoo+\Eoi)$, $b \equiv \tr(\Eoo+\Eio)$.  We want to make $\tr\{\Eoo\}$ as large as possible while keeping $\rho_A$ and $\rho_B$ fixed and having $\Eoi,\Eio,\Eii$ all still positive.  This corresponds to the measurement for which the particular outcomes $(0,0)$ are most probable.  
Assume first that $\Eoo=c R$ where $c$ is a constant and $0 < R < \hat{I}$. 
Note that 
\beq
\label{supp}
{\rm supp}(\Eoi)\cap {\rm supp}(\Eio)=\emptyset.
\eeq
Otherwise, we can always move the intersection part to $\Eoo$ to make the above statement true. 
Since
\beq
\begin{split}
\label{ine}
\Eoi \propto &\ \rho_A- \frac{c}{a}R  \geq 0 , \\
\Eio \propto &\ \rho_B- \frac{c}{b}R  \geq 0 , \\
%\Eii =& \id - a\rho_A - b\rho_B + \Eoo  \geq 0 .
\end{split}
\eeq
together with (\ref{supp}), implies that 
$$\frac{c}{a}R=\frac{c}{b}R=\Rhat.$$
Thus, the constant $c$ is chosen to equal to $a$ and $b$, and matrix $R=\Rhat$.
We have one last requirement $$\Eii = \id - c(\rho_A+\rho_B-\Rhat) \ge 0 .$$
From that, we can choose $c$ to be the largest value such that $\Eii \ge 0$.  
From the solution for $\Rhat$ in section 4, we see that
\begin{equation}
\label{valc}
(1/c) = \max_{\ket\psi, \bracket\psi\psi = 1} \frac{1}{2} \bra\psi \left( \rho_A + \rho_B + |\rho_A-\rho_B| \right) \ket\psi .
\end{equation}
The probability of $(0,0)$ outcome is then
\begin{equation}
p_{00}=\frac{\tr\{\Eoo\}}{D}=c\frac{\tr\{\Rhat\}}{D} .
\end{equation}

We now show that the probability of $(0,0)$ outcome will be maximized if $\tr\{\Rhat\}$ is as large as possible within the constraints given by (\ref{MR}). Consider a different positive operator
$\Rhat'$ that satisfies
\beq 
\begin{split}
\Rhat'-\Rhat\geq 0 , \\
\rho_A-\Rhat'\geq 0 , \\
\rho_B-\Rhat'\geq 0 .
\end{split}
\eeq
Defining $\Eoo=c\Rhat'$ retains the positivity of $\{\hat{E}_{ij}\}$, while the probability of $(0,0)$ outcome becomes
\begin{equation}
p_{00}' = \frac{\tr \Eoo}{D} = \frac{c\Rhat'}{D} = p_{00} + c\frac{\tr(\Rhat-\Rhat')}{D} \geq p_{00} .
\end{equation}
So to maximize the probability of the $(0,0)$ outcome, we must choose $\tr\{\Rhat\}$ as large as possible subject to the constraints of (\ref{MR}).  This concludes our proof.  
The combined density matrix will then be $\rho_{AB} = \Rhat/\tr\{\Rhat\}$, 
and the probability of the outcome will be
\begin{equation}
p_{00} = c \frac{K(\rho_A,\rho_B)}{D} ,
\end{equation}
where $c$ is given in (\ref{valc}).

We have shown that we can actually find the particular measurement such that when two parties pool their information, the probability of the joint outcome is maximized.  Note, however, that this required us to make very particular assumptions about the initial state (maximally mixed---the state of maximal ignorance) and the type of measurements that were done (positive measurement operators).  With a different initial state or more general measurements, this result need not have held.  Such assumptions must always be made to justify a particular choice of compatibility measure and information pooling procedure.

\section{Extension and relation to other compatibility measures}
\label{SDP_sec}

The measure of state compatibility in Sec.~\ref{SMR} was defined for two states, $\rho_A$ and $\rho_B$. This can be generalized to a set of $k$ states, $\mathcal{P}=\{\rho_i\}_{i=1}^{k}$, by defining the compatibility of the set $\mathcal{P}$, $K_{\textrm{BFM}}\left(\mathcal{P}\right)$, to be the value obtained from the following optimization:
\begin{align}
\begin{split}
\underset{R}{\textrm{Maximize: }}&\tr\left[R\right],\\ \label{k State BFM}
\textrm{Subject to: }& R\leq\rho_i, \ 1\leq i\leq k;\quad R \geq 0 .
\end{split}
\end{align}
While no simple formula is known for this measure for any number of states $k$, the optimizations in Eqs.~(\ref{MR}, \ref{k State BFM}) are examples of a general class of problems called {\it semidefinite programs} (SDPs) \cite{Vandenberghe1996, Watrous2011}. These are efficiently numerically solvable (e.g., using \cite{Lofberg2004, Sturm1999}).

Within the SDP formalism, for a given problem it is possible to construct a related dual problem (see Appendix \ref{SDP Appendix} for an outline of how to do so). For the SDP in Eq. (\ref{k State BFM}) this dual problem is
\begin{align}
\begin{split} \label{k State BFM Dual}
\underset{\mathcal{M}=\{M_i\}_{i=1}^{k}}{\textrm{Minimize: }}&\sum_{i=1}^{k}\tr\left[\rho_i M_i\right] ,\\
\textrm{Subject to: }&\sum_{i=1}^{k}M_i\geq \mathbb{I} ; \quad M_i\geq0,\  \forall i.
\end{split}
\end{align}
Any feasible solution to this dual SDP upper-bounds $K_{\textrm{BFM}}\left(\mathcal{P}\right)$ due to the property of weak duality that all SDPs obey. Furthermore, the two optimal solutions will be equal. (This can be shown using Slater's theorem.) Note that the set $\mathcal{M}=\{M_i\}_{i=1}^{k}$ resembles a measurement, except that it is required to sum to {\it more} than the identity in the SDP's constraints. It would be interesting if there was an operational interpretation of this.

\subsection{Post-Peierls (PP) Compatibility}

The SDPs given in Eqs.~(\ref{k State BFM}) and (\ref{k State BFM Dual}) are similar in form to an SDP found in \cite{Bandyopadhyay2014} where the task of {\it state exclusion} is addressed. State exclusion asks: given a system prepared in an unknown state from a given set $\mathcal{P}=\{\rho_i\}_{i=1}^{k}$, when is it possible to perform a measurement on the unknown state to conclusively rule out one of the preparations? If it is not possible, how small an error can one make when attempting to exclude a preparation by performing a measurement?

The task of state exclusion can be regarded as encapsulating another compatibility criteria, that of {\it Post-Peierls} (PP) compatibility \cite{Caves:2002ec}. For a set of states to be compatible in the PP sense, for any measurement that could be performed, there should exist at least one outcome that would occur with non-zero probability for all states in the set. Hence, if state exclusion is possible, the set of states is PP incompatible. How small an error it is possible to make when attempting to exclude a state gives a measure of the set's PP compatibility.

We can use the SDP from \cite{Bandyopadhyay2014} to define a measure  $K_{\textrm{PP}}\left(\mathcal{P}\right)$ of PP compatibility for a set $\mathcal{P}$, as the result of the following optimization:
\begin{dfn} The PP compatibility of a set $\mathcal{P}$, $K_{\textrm{PP}}\left(\mathcal{P}\right)$, is defined as:
\begin{align}
\begin{split}
\underset{N}{\textrm{Maximize: }}&\tr\left[N\right] ,\\
\textrm{Subject to: }&N\leq {\rho}_i, \ 1 \leq i\leq k; \quad N = N^\dagger.
\label{k State PP}
\end{split}
\end{align}
\end{dfn}
The associated dual to this SDP is:
\begin{align}
\begin{split}
\underset{\mathcal{M}=\left\{M_i\right\}_{i=1}^{k}}{\textrm{Minimize: }}&\sum_{i=1}^{k}\tr\left[{\rho}_i M_i\right] , \\
\textrm{Subject to: }&\sum_{i=1}^{k}M_i=\mathbb{I}; \quad M_i\geq0, \  \forall i.
\label{k State PP Dual}
\end{split}
\end{align}
By strong duality, the result of both optimizations Eqs.~(\ref{k State PP}) and (\ref{k State PP Dual}) will be the same.  Notice the similarity in form to Eqs.~(\ref{k State BFM}) and (\ref{k State PP}) on the one hand, and to Eqs.~(\ref{k State BFM Dual}) and (\ref{k State PP Dual}) on the other.

Caves \textit{et~al.}~show that if a set of states is BFM compatible, then it is also PP compatible \cite{Caves:2002ec}. This can readily be seen from the two measures of compatibility defined by Eqs.~(\ref{k State BFM}) and (\ref{k State PP}). If the set $\mathcal{P}$ is BFM compatible then $K_{\textrm{BFM}}\left(\mathcal{P}\right)>0$ and there exists a positive semidefinite $R$ satisfying the constraints of Eq.~(\ref{k State BFM}) such that $\tr[R]>0$. Being positive semidefinite implies that $R$ is Hermitian, so by setting $N=R$ we obtain an $N$ satisfying the constraints of Eq.~(\ref{k State PP}) such that $\tr[N]>0$. This implies that $K_{\textrm{PP}}\left(\mathcal{P}\right)>0$ and the set is also PP compatible. Thus, we have the following result:
\begin{theo}
$K_{\textrm{PP}}\left(\mathcal{P}\right)\geq K_{\textrm{BFM}}\left(\mathcal{P}\right)$.
\end{theo}

\subsection{Equal Support (ES) Compatibility}

A third compatibility measure for quantum states, stronger than both BFM and PP compatibility, is that of {\it equal support} (ES) \cite{Caves:2002ec}. A set of states $\mathcal{P}$ is ES compatible if and only if the states have the same support. Again, we can define a measure of ES compatibility $K_{\textrm{ES}}(\mathcal{P})$ by an SDP:
\begin{dfn} The ES compatibility of a set $\mathcal{P}$, $K_{\textrm{ES}}\left(\mathcal{P}\right)$, can be defined as:
\begin{align}
\begin{split}
\underset{\lambda}{\textrm{Maximize: }}&\lambda,\\
\textrm{Subject to: }&\lambda\sum_{j=1}^{k}\rho_j\leq\rho_i \ 1\leq i\leq k; \lambda\geq0.
\label{k State ES}
\end{split}
\end{align}
\end{dfn}

It can be shown (Appendix \ref{ES SDP Appendix}) that the dual SDP is:
\begin{align}
\begin{split}
\underset{\left\{\alpha_i\right\}_{i=1}^{D},\mathcal{M}=\left\{M_i\right\}_{i=1}^{k}}{\textrm{Minimize: }}&\sum_{i=1}^{k}\tr\left[{\rho}_i M_i\right], \\
\textrm{Subject to: }&\sum_{i=1}^{D}\alpha_i\geq1; \\
&\sum_{j=1}^{k}\rho_j \sum_{i=1}^{k} M_i \geq
\begin{pmatrix}
\alpha_1 & & \\
& \ddots &\\
& & \alpha_D
\end{pmatrix};\\
&\alpha_i\in\mathbb{R}, \alpha_i\geq0,  M_i\geq 0,\  \forall i.
\label{k State ES Dual}
\end{split}
\end{align}
Again, the primal and dual SDPs will give the same value, and will return 0 if the states in $\mathcal{P}$ are ES incompatible as they do not have equal support.

Caves \textit{et~al.}~show that if a set of states is ES compatible, then it is also BFM compatible. Equivalently, if the states are BFM incompatible, then they are also ES incompatible. This can be rederived by comparing the SDPs in Eqs.~(\ref{k State BFM Dual}) and (\ref{k State ES Dual}). If $\mathcal{P}$ is BFM incompatible, then there exists a set $\mathcal{M}$ that satisfies the constraints of Eq.~(\ref{k State BFM Dual}) such that:
\begin{equation}
\sum_{i=1}^{k} \tr\left[{\rho}_i M_i\right]=0.
\end{equation}
By picking the set $\left\{\alpha_i\right\}_{i=1}^{D}$ to be the eigenvalues of $\sum_{j=1}^{k}\rho_j$, $\left(\left\{\alpha_i\right\}_{i=1}^{D},\mathcal{M}\right)$ will be a feasible solution to the SDP in Eq.~(\ref{k State ES Dual}) that returns $K_{\textrm{ES}}=0$. Therefore, $\mathcal{P}$ is also ES incompatible. The fact that a feasible solution to the BFM dual SDP can be used as the basis for a feasible solution to the ES dual gives:
\begin{theo}
$K_{\textrm{BFM}}\left(\mathcal{P}\right)\geq K_{\textrm{ES}}\left(\mathcal{P}\right)$.
\end{theo}

\section{Conclusions}
\label{sec7}

The problem of state compatibility is rather subtle, because state assignments do not perfectly reflect the information used to create them.  Different measures of compatibility (and different methods of pooling information) may make sense for different ways that Alice and Bob may have acquired their information.  It is certainly impossible to define a measure of compatibility without some assumption about what type of information was used to produce the state assignments.  By contrast, the qualitative criterion of \cite{Brun:2002eo} only requires that the states {\it might} describe the same system.

Looking at different methods of acquiring information, however, remains an interesting question.  Within a particular assumption about information gathering, it may make sense to define measures of compatibility and methods of pooling states, and these might be useful in practice.  We defined one such measure---which is the solution of a semidefinite program---and one such pooling method, which have an interpretation based on the assumption that the observers' state assignments derive from information acquired by measurements of a particular form.  There is still considerable room for development in this area.

Moreover, as discussed in Sec.~\ref{SDP_sec}, there are other quite different approaches to the problem of state compatibility \cite{Caves:2002ec}, based on the existence of measurements that discriminate between pairs of states; these approaches lead to an entire hierarchy of compatibility conditions.  We have shown that for two of these criteria (PP and ES compatibility), we can define measures of compatibility based on SDPs, just as for BFM compatibility.  It would be interesting to determine whether compatibility measures for the other definitions given in \cite{Caves:2002ec} (Weak and Weak$^\prime$) also can be formulated as SDPs.  The status of these other idea of compatibility leaves much to be explored, and may be the key to understanding compatibility and the ability (or inability) to achieve consensus between observers with different priors.

\section*{Acknowledgments}
We would like to thank Alexei Kitaev, Michael Nielsen, and R\"udiger Schack, for very useful conversations, and Robin Blume-Kohout, Carl Caves, Jennifer Dodd, Jerry Finkelstein, Chris Fuchs, Kurt Jacobs, David Mermin, David Poulin, and Howard Wiseman for their interactions and feedback.  This work was supported in part by NSF CAREER Grant No.~CCF-0448658. MH is currently supported by an ARC Future Fellowship under Grant FT140100574. 
%\end{acknowledgments}

\clearpage
\appendix
\section{Semidefinite Programs} \label{SDP Appendix}

Here we give the form of a semidefinite program and the relation between the primal and dual problems as formulated in \cite{Watrous2011}. An SDP is formed of three elements, $\{A,B,\Phi\}$. $A$ and $B$ are Hermitian matrices and $\Phi$ is a Hermicity preserving superoperator.

Using these, we define the primal problem to be:
\begin{align}
\begin{split} \label{Prime}
\underset{X}{\textrm{Maximize:}} & \quad \alpha=\tr[AX],\\
\textrm{Subject to:} & \quad \Phi(X)\leq B; \quad X\geq0.
\end{split}
\end{align}
The related dual is then given by:
\begin{align}
\begin{split} \label{Dual}
\underset{Y}{\textrm{Minimize:}} & \quad \beta=\tr[BY],\\
\textrm{Subject to}:& \quad \Phi^{*}(Y)\geq A; \quad Y\geq0.
\end{split}
\end{align}
Here $\Phi^*$ is the dual map to $\Phi$, given by:
\beq
\tr[Y\Phi(X)]=\tr[X\Phi^{*}(Y)]. \label{Phi* equation}
\eeq

\subsection{ES Compatibility SDP} \label{ES SDP Appendix}

Here we show that the SDP given in Eq.~(\ref{k State ES Dual}) is the dual to that defined in Eq.~(\ref{k State ES}).  First we rewrite Eq.~(\ref{k State ES}) so that it has the same structure as Eq.~(\ref{Prime}). This leads to:

\begin{align}
\begin{split}
\underset{\lambda,\left\{\lambda_i\right\}_{i=1}^{D}}
{\textrm{Maximize:}}& \\ 
\alpha&=\tr\left[
\begin{pmatrix}
\lambda & & & \\
& \lambda_1 & &\\
& & \ddots &\\
& & & \lambda_D
\end{pmatrix}
\begin{pmatrix}
1 & & & \\
& 0 & & \\
& & \ddots &\\
& & & 0
\end{pmatrix}
\right].\\
\textrm{Subject to:}& \quad\lambda-\lambda_i \leq 0, \quad\forall i; \\
& \quad 
\begin{pmatrix}
\lambda_1 & &\\
& \ddots &\\
& & \lambda_D
\end{pmatrix}
\sum_{j=1}^{k} \rho_j \leq \rho_i, \quad\forall i;\\
&\quad\lambda\geq0; \quad \lambda_i\geq0, \  \forall i.
\end{split}
\label{ES SDP rewrite}
\end{align}

%\begin{widetext}
%\begin{align}
%\begin{split}
%\underset{
%\lambda,\left\{\lambda_i\right\}_{i=1}^{d}
%}{\textrm{Maximize:}} \quad\quad \alpha&=\tr\left[
%\begin{pmatrix}
%\lambda & & & \\
%& \lambda_1 & &\\
%& & \ddots &\\
%& & & \lambda_D
%\end{pmatrix}
%\begin{pmatrix}
%1 & & & \\
%& 0 & & \\
%& & \ddots &\\
%& & & 0
%\end{pmatrix}
%\right].\\
%\textrm{Subject to:} \quad\quad \lambda-\lambda_i &\leq 0, \quad\forall i; \quad 
%\begin{pmatrix}
%\lambda_1 & &\\
%& \ddots &\\
%& & \lambda_D
%\end{pmatrix}
%\sum_{j=1}^{k} \rho_j \leq \rho_i, \quad\forall i;\\
%\quad\lambda&\geq0; \quad \lambda_i\geq0, \  \forall i.
%\end{split}
%\label{ES SDP rewrite}
%\end{align}
%\end{widetext}

Comparing Eq.~(\ref{ES SDP rewrite}) with Eq.~(\ref{Prime}), we see that:
\begin{itemize}
\item $A$ is a $D+1$ by $D+1$ matrix:
\begin{equation}
A=\begin{pmatrix}
1 & & & \\
& 0 & & \\
& & \ddots &\\
& & & 0
\end{pmatrix}.
\end{equation}
\item $B$ is a $D(k+1)$ by $D(k+1)$ matrix where the first $D$ entries on the diagonal are 0, and the remaining matrix is block diagonal with the blocks given by $\rho_i$:
\begin{equation}
B=\begin{pmatrix}
0 & & & & &\\
& \ddots & & & &\\
& & 0 & & &\\
& & & \rho_1 & &\\
& & & & \ddots &\\
& & & & & \rho_k
\end{pmatrix}.
\end{equation}
\item $X$, the variable matrix, is a $D+1$ by $D+1$ matrix:
\begin{equation}
X=\begin{pmatrix}
\lambda & & & \\
& \lambda_1 & & \\
& & \ddots &\\
& & & \lambda_D
\end{pmatrix}.
\end{equation}
\item $Y$ is a $D(k+1)$ by $D(k+1)$ matrix whose first $D$ entries on the diagonal we label by $\alpha_i$, and the remaining block diagonal with the elements we denote by $M_i$:
\begin{equation}
Y=\begin{pmatrix}
\alpha_1 & & & & &\\
& \ddots & & & &\\
& & \alpha_D & & &\\
& & & M_1 & &\\
& & & & \ddots &\\
& & & & & M_k
\end{pmatrix}.
\end{equation}
\item The map $\Phi$ is given by:
\end{itemize}
\begin{widetext}
\begin{equation}
\Phi\left(X\right)=
\begin{pmatrix}
\lambda-\lambda_1 & & & & &\\
& \ddots & & & &\\
& & \lambda-\lambda_D & & & \\
& & &
\begin{pmatrix}
\lambda_1 & & \\
& \ddots &\\
& & \lambda_D
\end{pmatrix}\sum_{i=1}^{k} \rho_i
& & \\
& & & & \ddots &\\
& & & & & \begin{pmatrix}
\lambda_1 & & \\
& \ddots &\\
& & \lambda_D
\end{pmatrix}\sum_{i=1}^{k} \rho_i
\end{pmatrix}.
\end{equation}

Using Eq.~(\ref{Phi* equation}), we see that $\Phi^*$ must satisfy:
\begin{equation}
\begin{split}
\sum_{i=1}^{D}\alpha_i\left(\lambda-\lambda_i\right)+&\sum_{i=1}^{k}\tr\left[M_i\begin{pmatrix}
\lambda_1 & & \\
& \ddots &\\
& & \lambda_D
\end{pmatrix}\sum_{j=1}^{k} \rho_j\right]\\
=&
\tr\left[
\begin{pmatrix}
\lambda & & & \\
& \lambda_1 & & \\
& & \ddots &\\
& & & \lambda_D
\end{pmatrix}\Phi^*\left(
\begin{pmatrix}
\alpha_1 & & & & &\\
& \ddots & & & &\\
& & \alpha_D & & &\\
& & & M_1 & &\\
& & & & \ddots &\\
& & & & & M_k
\end{pmatrix}
\right)
\right],
\end{split}
\end{equation}
and hence $\Phi^*(Y)$ produces a $D+1$ by $D+1$ matrix:
\begin{equation}
\Phi^*\left(Y\right)
=\begin{pmatrix}
\sum_{i=1}^{D}\alpha_i &\\
& \begin{pmatrix}
-\alpha_1 & &\\
& \ddots &\\
& & -\alpha_D
\end{pmatrix}+\sum_{i=1}^{k} \rho_i \sum_{j=1}^{k}M_i
\end{pmatrix}.
\end{equation}

If we now substitute these elements into Eq.~(\ref{Dual}), we obtain Eq.~(\ref{k State ES Dual}). 
\end{widetext}


\begin{thebibliography}{99}

\bibitem{Brun:2002eo} T. Brun, J. Finkelstein, and N. Mermin, Phys. Rev. A 65, 032315 (2002).
\bibitem{Jacobs:2002cc} K. Jacobs, Quantum Inf Process 1, 73 (2002).
\bibitem{Caves:2002ec} C. Caves, C. Fuchs, and R. Schack, Phys. Rev. A 66, 062111 (2002).
\bibitem{Poulin:2003he} D. Poulin and R. Blume-Kohout, Phys. Rev. A 67, 010101 (2003).
\bibitem{Herbut:2004vc} F. Herbut, J. Phys. A: Math. Gen. 37, 5243 (2004).
\bibitem{Jacobs:829374} K. Jacobs, Phys. Rev. A 72, 044101 (2005). 
\bibitem{Spekkens:2007wn} R. W. Spekkens and H. M. Wiseman, Phys. Rev. A 75, 042104 (2007).
\bibitem{Leifer:2011ug} M. S. Leifer and R. W. Spekkens, arXiv:1110.1085 (2011).
\bibitem{Leifer:2011wq} M. S. Leifer and R. W. Spekkens, arXiv:1107.5849, (2011).
\bibitem{Jaynes} E. T. Jaynes, Phys. Rev {\bf 106}, 620 (1957); E. T. Jaynes, Phys. Rev. {\bf 108}, 171 (1957).
\bibitem{Kitaev} A.~Kitaev, private communication.
\bibitem{Nielsen} M. A.~Nielsen, private communication.
\bibitem{NielsenChuang} M. A.~Nielsen and I. L.~Chuang {\sl Quantum Computation and Quantum Information} (Cambridge University Press, Cambridge, 2000).
\bibitem{Vandenberghe1996} L. Vandenberghe and S. Boyd, SIAM review 38, 49 (1996).
\bibitem{Watrous2011} J. Watrous, Lecture notes, CS 766/QIC 820 Theory of Quantum Information, University of Waterloo. See lecture 7, Semidefinite programming, and Lecture 8, Semidefinite Programs for Fidelity and Optimal Measurements (Fall, 2011).
\bibitem{Lofberg2004} J. L\"ofberg, in Computer Aided Control Systems Design, 2004 IEEE International Symposium on (IEEE, 2004) pp. 284?289.
\bibitem{Sturm1999} J. F. Sturm, Optimization methods and software 11, 625 (1999).
\bibitem{Bandyopadhyay2014} S. Bandyopadhyay, R. Jain, J. Oppenheim, and C. Perry, Phys. Rev. A 89, 022336 (2014).
\end{thebibliography}
\end{document}